\documentclass[letterpaper,twocolumn,prl,aps,superscriptaddress,floatfix]{revtex4-2}
\usepackage{amsmath}
\usepackage{amssymb}
\usepackage{graphicx}
\usepackage{esint}
\usepackage{times}
\usepackage[markup=underlined]{changes}

\usepackage[unicode=true,
bookmarks=true,bookmarksnumbered=false,bookmarksopen=false,
breaklinks=false,pdfborder={0 0 1},backref=false,colorlinks=true]
{hyperref}
\hypersetup{
    linkcolor=magenta, urlcolor=blue, citecolor=blue, pdfstartview={FitH}, hyperfootnotes=false, unicode=true}

\makeatletter

\pdfpageheight\paperheight
\pdfpagewidth\paperwidth

\@ifundefined{textcolor}{}
{%
    \definecolor{BLACK}{gray}{0}
    \definecolor{WHITE}{gray}{1}
    \definecolor{RED}{rgb}{1,0,0}
    \definecolor{GREEN}{rgb}{0,1,0}
    \definecolor{BLUE}{rgb}{0,0,1}
    \definecolor{CYAN}{cmyk}{1,0,0,0}
    \definecolor{MAGENTA}{cmyk}{0,1,0,0}
    \definecolor{YELLOW}{cmyk}{0,0,1,0}
}

\usepackage{xcolor}
\usepackage{soul}

\usepackage[mathlines]{lineno}
\modulolinenumbers[5] 

\definecolor{blue}{rgb}{0,0,1}
\definecolor{red}{rgb}{1,0,0}
\definecolor{green}{rgb}{0,1,0}

\usepackage{soul}

\begin{document}

\title{A re-entrant chip-free-space photonic interface for telecom-to-Rubidium spectroscopy}

\author{Jia-Lin Chen}
\affiliation{Laboratory of Quantum Information, University of Science and Technology of China, Hefei 230026, China.}
\affiliation{Anhui Province Key Laboratory of Quantum Network, University of Science and Technology of China, Hefei 230026, China.}
\affiliation{CAS Center for Excellence in Quantum Information and Quantum Physics, University of Science and Technology of China, Hefei 230026, China.}

\author{Ruixin Zhou}
\affiliation{School of Information Science and Technology, ShanghaiTech University, Shanghai 201210, China.}
\affiliation{State Key Laboratory of Quantum Functional Materials, ShanghaiTech University, Shanghai 201210, China.}

\author{Deng-Hong Liu}
\affiliation{Laboratory of Quantum Information, University of Science and Technology of China, Hefei 230026, China.}
\affiliation{Anhui Province Key Laboratory of Quantum Network, University of Science and Technology of China, Hefei 230026, China.}
\affiliation{CAS Center for Excellence in Quantum Information and Quantum Physics, University of Science and Technology of China, Hefei 230026, China.}

\author{You-Long Fan}
\affiliation{Laboratory of Quantum Information, University of Science and Technology of China, Hefei 230026, China.}
\affiliation{Anhui Province Key Laboratory of Quantum Network, University of Science and Technology of China, Hefei 230026, China.}
\affiliation{CAS Center for Excellence in Quantum Information and Quantum Physics, University of Science and Technology of China, Hefei 230026, China.}

\author{Zhu-Bo Wang}
\affiliation{Laboratory of Quantum Information, University of Science and Technology of China, Hefei 230026, China.}
\affiliation{Anhui Province Key Laboratory of Quantum Network, University of Science and Technology of China, Hefei 230026, China.}
\affiliation{CAS Center for Excellence in Quantum Information and Quantum Physics, University of Science and Technology of China, Hefei 230026, China.}
\affiliation{Hefei National Laboratory, University of Science and Technology of China, Hefei 230088, China.}

\author{Min Chen}
\affiliation{Laboratory of Quantum Information, University of Science and Technology of China, Hefei 230026, China.}
\affiliation{Anhui Province Key Laboratory of Quantum Network, University of Science and Technology of China, Hefei 230026, China.}
\affiliation{CAS Center for Excellence in Quantum Information and Quantum Physics, University of Science and Technology of China, Hefei 230026, China.}

\author{Xiang Fang}
\affiliation{Laboratory of Quantum Information, University of Science and Technology of China, Hefei 230026, China.}
\affiliation{Anhui Province Key Laboratory of Quantum Network, University of Science and Technology of China, Hefei 230026, China.}
\affiliation{CAS Center for Excellence in Quantum Information and Quantum Physics, University of Science and Technology of China, Hefei 230026, China.}

\author{Jia-Qi Wang}
\affiliation{Laboratory of Quantum Information, University of Science and Technology of China, Hefei 230026, China.}
\affiliation{Anhui Province Key Laboratory of Quantum Network, University of Science and Technology of China, Hefei 230026, China.}
\affiliation{CAS Center for Excellence in Quantum Information and Quantum Physics, University of Science and Technology of China, Hefei 230026, China.}

\author{Zheng-Fu~Han}
\affiliation{Laboratory of Quantum Information, University of Science and Technology of China, Hefei 230026, China.}
\affiliation{Anhui Province Key Laboratory of Quantum Network, University of Science and Technology of China, Hefei 230026, China.}
\affiliation{CAS Center for Excellence in Quantum Information and Quantum Physics, University of Science and Technology of China, Hefei 230026, China.}
\affiliation{Hefei National Laboratory, University of Science and Technology of China, Hefei 230088, China.}

\author{Guang-Can Guo}
\affiliation{Laboratory of Quantum Information, University of Science and Technology of China, Hefei 230026, China.}
\affiliation{Anhui Province Key Laboratory of Quantum Network, University of Science and Technology of China, Hefei 230026, China.}
\affiliation{CAS Center for Excellence in Quantum Information and Quantum Physics, University of Science and Technology of China, Hefei 230026, China.}
\affiliation{Hefei National Laboratory, University of Science and Technology of China, Hefei 230088, China.}

\author{Ai-Ping Liu}
\affiliation{School of Communications and Information Engineering, Nanjing University of Posts and Telecommunications, Nanjing 210003, China.}

\author{Pengfei Wang}
\affiliation{Innovation Academy for Precision Measurement Science and Technology, Chinese Academy of Sciences, Wuhan 430071, China.}

\author{Xiaochi Liu}
\email{liuxc@apm.ac.cn}
\affiliation{Innovation Academy for Precision Measurement Science and Technology, Chinese Academy of Sciences, Wuhan 430071, China.}

\author{Juanjuan~Lu}
\email{lujj2@shanghaitech.edu.cn}
\affiliation{School of Information Science and Technology, ShanghaiTech University, Shanghai 201210, China.}
\affiliation{State Key Laboratory of Quantum Functional Materials, ShanghaiTech University, Shanghai 201210, China.}

\author{Wei Chen}
\email{weich@ustc.edu.cn}
\affiliation{Laboratory of Quantum Information, University of Science and Technology of China, Hefei 230026, China.}
\affiliation{Anhui Province Key Laboratory of Quantum Network, University of Science and Technology of China, Hefei 230026, China.}
\affiliation{CAS Center for Excellence in Quantum Information and Quantum Physics, University of Science and Technology of China, Hefei 230026, China.}
\affiliation{Hefei National Laboratory, University of Science and Technology of China, Hefei 230088, China.}

\author{Chang-Ling Zou}
\affiliation{Laboratory of Quantum Information, University of Science and Technology of China, Hefei 230026, China.}
\affiliation{Anhui Province Key Laboratory of Quantum Network, University of Science and Technology of China, Hefei 230026, China.}
\affiliation{CAS Center for Excellence in Quantum Information and Quantum Physics, University of Science and Technology of China, Hefei 230026, China.}
\affiliation{Hefei National Laboratory, University of Science and Technology of China, Hefei 230088, China.}

\date{\today}

\begin{abstract}
\textbf{Photonic integrated circuits (PICs) generate, route, and process light with high efficiency, scalability, and functional density on a single chip. Yet the tightly confined on-chip modes can not easily access or effectively interact with atomic vapors, fluids, gain media, and  biological samples. Existing approaches require bringing the medium onto the chip or into a weak, tightly confined evanescent field, which restricts the interaction volume and the range of accessible media. Here, we demonstrate a re-entrant chip-free-space interface in which a thin-film lithium niobate circuit frequency-doubles telecom light, emits the 780~nm field through a Rubidium vapor cell, and recollects the reflected probe on the same chip. This emit-interact-recollect loop resolves the saturated absorption spectrum and stabilizes the telecom laser to within $\pm 280$~kHz over 2 hours. Our study paves an route to embed external media into PICs through the re-entrant photonic interface.}
\end{abstract}

\maketitle

\noindent \textbf{\large{}Introduction}{\large\par}
\noindent Photonic integrated circuits (PICs) have become a foundational technology for optical communication, computing, sensing, and metrology, because their planar, on-chip integration offers large-scale manufacturability, high integration density, and the ability to generate, route, modulate, filter, and detect light with high  efficiency on a single chip~\cite{Pelucchi2022,Shekhar2024,stojanovic_Monolithic_2018,zhu_Integrated_2021,liu_Aluminum_2023}. The PIC benefits from tight optical field confinement in micron-scale waveguides, which enhances nonlinear interactions and packs many functions into a compact footprint~\cite{wang_Optical_2024a,hu_Integrated_2025}. A broad and growing class of applications, however, cannot be served by such two-dimensional integration alone, but instead demands that a free-space optical beam be merged with the PIC. The reasons are physical rather than incidental. A dilute gas or vapor demands an interaction volume far larger than any evanescent field~\cite{vanier_Quantum_2024}. A gain medium must be driven below its damage or saturation threshold, which calls for an expanded free-space mode rather than a tightly confined one~\cite{ristau_LaserInduced_2014}. Biological and condensed samples have sizes and shapes that cannot be patterned into a waveguide~\cite{wilson_Optical_2016}. Additionally, many functional materials are difficult to process on chip or have refractive indices incompatible with the guiding layer~\cite{Tran2022,Baets2023,Wang2023}. In all of these cases the on-chip processing is highly desirable, yet the essential light-matter interaction must take place in a free-space beam that a confined guided mode cannot form.

\begin{figure*}
    \centering
    \includegraphics[scale=0.9]{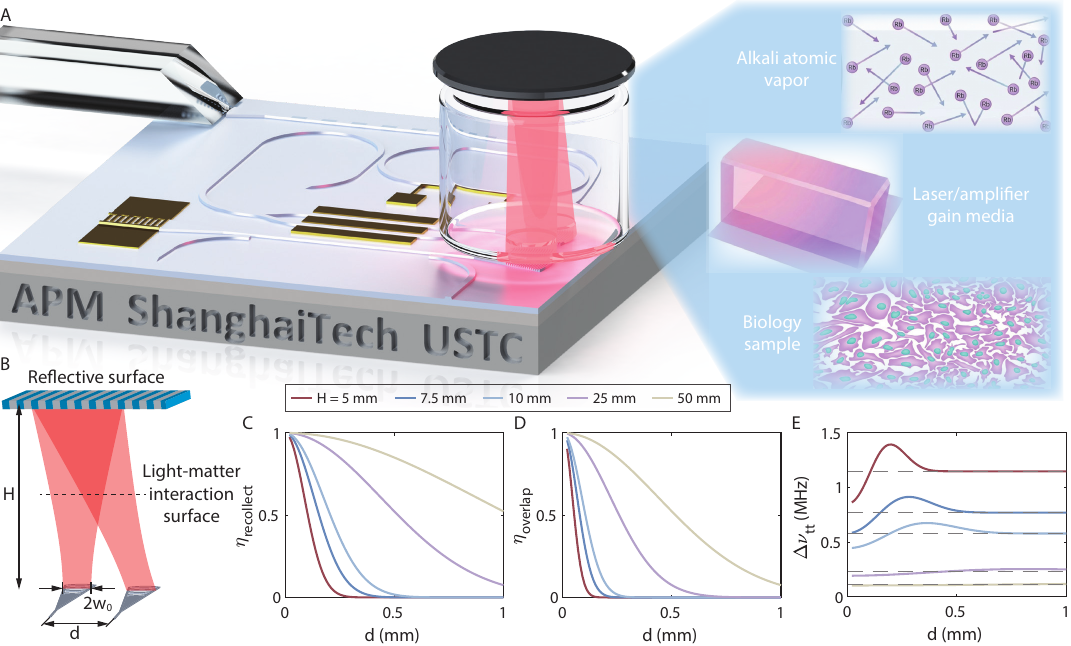}
    \caption{\textbf{The re-entrant chip-Cfree-space photonic interface.} (A) Schematic of a photonic integrated circuit co-integrated with an external medium, so that on-chip light is emitted into the medium and recollected on the same chip, making the medium a re-entrant functional element of the circuit. The external medium may be an alkali atomic vapor, an optical gain medium, or a biological analyte, among others. (B) Geometry of the emit-interact-recollect loop, defined by the emission and recollection grating couplers (GCs) of width $2w_0$ separated by a pitch $d$, and a reflective surface at height $H$ that redirects the emitted beam to the recollection GC. (C) Recollection efficiency $\eta_{\mathrm{recollect}}$ and (D) emission recollection beam overlap $\eta_{\mathrm{overlap}}$, both of which improve as the GC pitch is reduced, showing that a more compact interface remains feasible. (E) Transit-time broadening $\Delta\nu_{\mathrm{tt}}$, which stays about one order of magnitude below the intrinsic $^{87}$Rb linewidth at temperature $T=$300~K, so that spectral resolution is preserved.
    }
    \label{fig1}
\end{figure*}

\begin{figure*}
    \centering
    \includegraphics{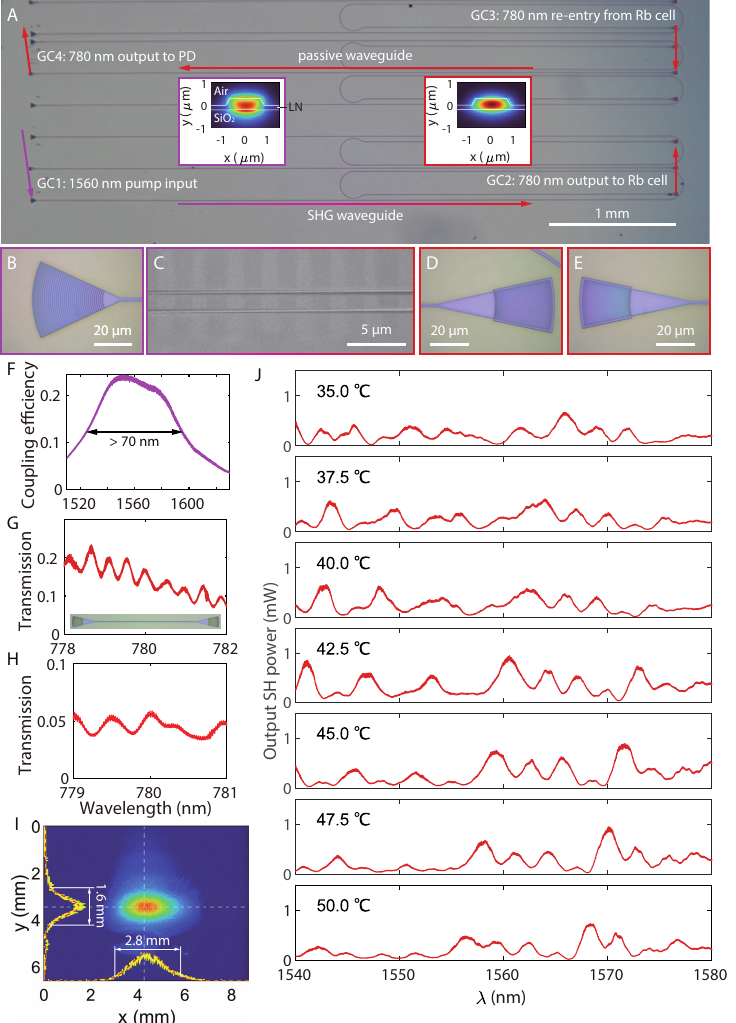}
    \caption{\textbf{The TFLN PPLN chip.} (A) Microscopic photo of the PPLN chip for second harmonic generation. The two insets are TM basic modes for 1560 nm and 780 nm light propagating in the waveguide, which are involved in the SHG process. The light goes through four GCs in the whole loop, which are noted as GC1, GC2, GC3 and GC4 in sequence. The arrows, frames and curves in magenta and red colors correspond to telecom band and visible band, respectively. (B) Telecom-band GC. (C) Polarized domain of the periodically poled waveguide captured by scanning electron microscope. (D) and (E) are visible-band GC for emitting to Rb cell and for coupling with fiber. (F) Telecom-band GC coupling efficiency spectrum. (G) Transmission spectrum of a 250~$\mu$m waveguide with visible-band GCs. (H) Transmission spectrum of the passive waveguide. (I) The emitted light beam 30~mm away from the visible-band GC. (J) Second harmonic generation spectrum under different temperature.}
    \label{fig2}
\end{figure*}

Several approaches have been developed to bridge this gap. The most direct way sends light off chip through optical fibers to a separate free-space apparatus, but this adds coupling and packaging at both the chip-fiber and fiber-instrument interfaces, enlarging the footprint and degrading stability, so that the very compactness and integration offered by the chip are lost~\cite{xie_Infrared_2019}. Alternatively, although the accessible media that compatible with the chip surface are limited, we can bring the them onto the chip and lets it interact with the weak evanescent field of a waveguide. In the atom-cladding waveguide schemes, for instance, the interaction is confined to a thin surface layer for coupling optical near-fields, which limits the interaction volume and induces severe transit-time broadening \cite{yang_Atomic_2007,stern_Nanoscale_2013,Zektzer2021,shrestha_Enabling_2026,zhou_Trapped_2024,xu_Dynamics_2025}. A potentially more scalable approach introduces an on-chip waveguide-mode to free-space beam converter, which launches a probe beam to interact with the target medium, while the interacted light is collected off chip on an external detector and never returns to the PIC~\cite{Newman2019,hummon_Photonic_2018,Sebbag2021}. Recently, Yulaev \textit{ et al.} took a first step toward closing this loop, using a single metasurface to emit a beam vertically through a vapor cell and recollect the reflected light on the same chip~\cite{yulaev_Chipscale_2024}. However, this shared-aperture configuration introduces unwanted back-reflection and interference in the detected signal, and its integration with other on-chip functional devices remains elusive.

Here, we introduce and demonstrate a general re-entrant photonic interface that incorporates free-space light-matter interaction into PICs through an emit-interact-recollect loop. On-chip light is emitted into free space by a grating coupler (GC), interacts with an external medium held above the chip, is returned by a reflector, and is recollected by a second grating coupler on the same chip for further on-chip processing. Unlike a shared-aperture geometry, the spatially separated emitting and recollecting GCs suppress direct back-reflection and stray-light interference, while turning the GC into a bidirectional chip-free-space port that embeds an otherwise non-integrable medium directly into the PIC. We show that the loop closes at the millimeter scale with relaxed alignment tolerances, and that the finite free-space beam keeps transit-time broadening about an order of magnitude below the intrinsic atomic linewidth, a regime inaccessible to evanescent schemes. As a proof of principle, we co-integrate a thin-film lithium niobate frequency-conversion chip with a Rubidium vapor cell: the chip frequency-doubles telecom light to 780 nm, emits it through the cell, and recollects the retro-reflected probe, resolving the $^{87}$Rb D2 saturated absorption spectrum (SAS) and locking the telecom laser to the atomic resonance within $\pm 280$ kHz over 2 hours, with no free-space frequency-conversion optics. More broadly, because each external medium can be held in a self-contained, millimeter-scale module that couples to the chip through a single standardized grating-coupler-reflector-grating-coupler footprint, the re-entrant interface establishes a general and scalable route to heterogeneously integrate atoms, gain media, fluids, and biological samples with PICs.

\smallskip{}
\noindent \textbf{\large{}Results}{\large\par}
\noindent Figure~\ref{fig1}A illustrates the central idea of the re-entrant chip-free-space interface. In a conventional PIC, light is generated, routed, and processed within tightly confined, micron-scale on-chip modes, and any interaction with an external medium takes place off chip and terminates in a free-space detector, so the interacted light never returns to the circuit \cite{hummon_Photonic_2018}. This either restricts the medium to specific types that can be brought onto the chip or requires the media surrounding the waveguide to interact with the weak evanescent field, limiting the interaction volume and the range of accessible media. The most natural port between an on-chip waveguide mode and a free-space beam is a GC, which routinely emits guided light into free space but is conventionally used only as a one-way output. Actually, closing the loop requires almost nothing more: if the emitted beam interacts with a medium held in a thin module of finite height above the chip and is then sent back by a reflector, whether a simple mirror, a designed reflection grating, or a metasurface, a second GC on the same chip suffices to recollect it. The separated GCs provide an intrinsic spatial-filtering function that suppresses stray light in the transmitted signal, avoiding interference in the detected signal~\cite{yulaev_Chipscale_2024}. A single emitting coupler and a single detector would be sufficient to probe the medium, but only recollection on chip returns the interacted light to the integrated circuit and enable further on-chip processing. The second coupler is therefore what is necessary to make the light path re-entrant. This minimal GC-mirror-GC unit is also sufficient, since it simultaneously delivers light to the medium and couples the returning signal back for further on-chip processing. The external medium, which may be an atomic vapor, a gain medium, or a biological analyte among others, thereby becomes a re-entrant functional element embedded within the on-chip processing chain rather than a terminal stage, giving the architecture access to media and interaction geometries that confined waveguide modes cannot reach.

To show that this architecture supports a complete functional system, we demonstrate a representative example an on-chip cycle of frequency doubling, emission, saturated absorption in a vapor, and recollection. Realizing the re-entrant requirement imposes constraints beyond the on-chip-waveguide-to-free-space conversion efficiency of a single coupler. The emitted and recollected beams should be spatially matched so that the returning field couples back into the recollection GC, and the forward (pump) and backward (probe) beams must overlap within the medium for SAS. Figure~\ref{fig1}B shows a two-dimensional model of the simplest realization: two identical GCs, each of width $2w_0$ and separated by a gap $d$, emit and recollect a Gaussian beam whose waist sits at the chip surface, while a reflective surface at height $H$, assumed to provide customized phase compensation, redirects the emitted beam onto the recollection coupler. We evaluate the beam overlap using an equivalent reciprocal picture in which both couplers emit virtual Gaussian beams, and compute the overlap of their field amplitudes at the reflective surface and, for the interaction, at the representative plane $H/2$ (see Supplementary Information for more details). We consider GC design with typical width of 20~$\mu$m and reflective surface height of millimeter order.

Figures~\ref{fig1}C and~\ref{fig1}D show, respectively, the recollection efficiency $\eta_{\mathrm{recollect}}$ and the emission-recollection beam overlap $\eta_{\mathrm{overlap}}$ as functions of the geometry. Making the system more compact reduces both quantities, but this reduction can be compensated by decreasing the grating-coupler pitch, so that a compact millimeter-scale interface remains feasible without stringent parameter control. For reflective surface height down to 5~mm, a GC pitch of 50~$\mu$m, which is easy to realize on chip, can still maintain $\eta_{\text{recollect}}$ at 85.1\% and $\eta_{\text{overlap}}$ at 52.5\%. A distinctive feature of the re-entrant geometry is that the finite beam size at the interaction height introduces transit-time broadening of the resonance linewidth, a physical regime that on-chip waveguides, with their micron-scale confined modes and correspondingly short atomic transit times, cannot access. As shown in Fig.~\ref{fig1}E, a smaller reflective-surface height $H$ yields a smaller beam and hence stronger broadening. As the emission-recollection overlap decreases, the broadening evolves smoothly toward the weak-saturation limit {$\Delta\nu_{\mathrm{tt}}\approx \frac{2\sqrt{2\ln 2}}{\pi}\sqrt{\frac{k_B T}{\pi m}}\,\frac{1}{w(H/2)}$}. Crucially, even for $H = 5$~mm the transit-time broadening remains at about 1~MHz level, well below the intrinsic $^{87}$Rb linewidth (6~MHz), so the free-space interaction preserves the atomic spectral resolution while the loop is closed on chip, an advantage that is intrinsic to the millimeter-scale free-space span of the re-entrant architecture and unavailable to purely on-chip evanescent schemes.

\begin{figure*}
    \centering
    \includegraphics{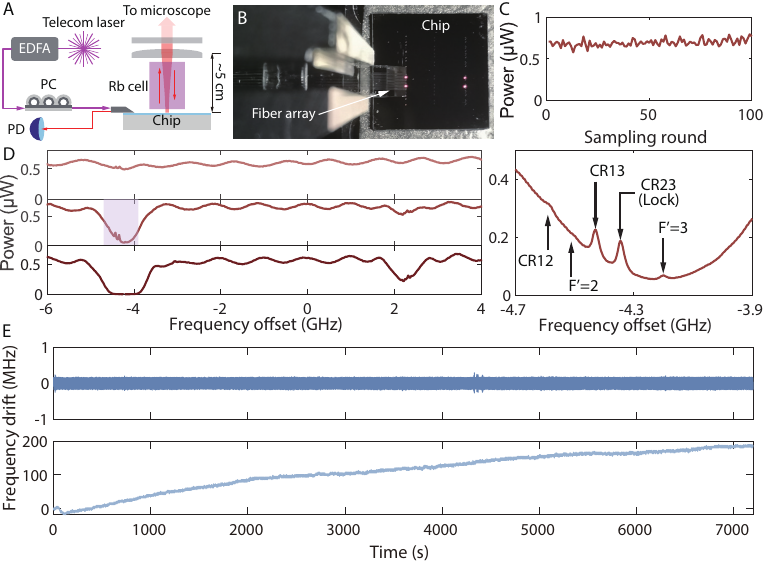}
    \caption{\textbf{Atomic spectrum interrogation and frequency locking.}  (A) Experiment setup for saturation absorption spectrum (SAS) interrogation. EDFA: erbium-doped fiber amplifier; PC: polarization controller; PD: photodetector. (B) Plane view of the chip, coupled with fiber array for light input and output. In this picture, 780~nm light is coupled into two passive waveguides from two channels of the fiber array respectively to check the alignment. In spectrum interrogation and frequency locking experiment, the optical input and output is instead as shown in Fig.~\ref{fig2}. (C) The probe power fluctuation detected by PD under sampling of 100 rounds at off-resonance frequency. (D) $^{87}$Rb $D_2$ line SAS under different cell temperature. The right inset: zooms in view of the SAS under 48.0$^{\circ}$C, revealing transition features of $5^2S_{1/2},F=2$ gound state. The feature peaks are labeled either with the total angular momentum quantum number $F'$ of the transition final state or two final states of the cross-resonances (CR). (E) Telecom-band laser frequency drift over 2~h when frequency locking is on (upper) and off (lower).}
    \label{fig3}
\end{figure*}

We experimentally implement such re-entrant chip-free-space system by co-integrate a TFLN frequency conversion chip with a Rb atomic frequency reference system and demonstrate their collaborative functioning. Figure~\ref{fig2}A shows the chip, which performs two roles within the re-entrant loop: it converts telecom-band light to the visible and emits it toward the Rb cell, and it recollects the probe light returning from the cell. These two roles are carried out by two waveguides on the same z-cut TFLN platform (500~nm TFLN on a 2~$\mu$m buried silica layer on silicon): an second harmonic genreation (SHG) waveguide that generates and emits the 780~nm field, and a passive waveguide that recollects and routes the returned light. Both are rib waveguides with 1.35~$\mu$m width, 60$^{\circ}$ sidewall angle, and 300~nm etch depth into the 500~nm film~\cite{yang_Nonlinear_2023}, a geometry chosen to balance the SHG efficiency against the GC efficiency. To characterize the passive waveguide independently, we measure on-chip micro-ring resonators (not shown) and extract intrinsic quality factors of $1.76\times 10^5$ at 1560~nm and $3.79\times 10^4$ at 780~nm, corresponding to propagation losses of 1.71~dB/cm and 4.37~dB/cm, respectively.

Now, we describe the optical path for the system.

(i) \textit{Input and SHG}. As shown in the left inset of Fig.~\ref{fig2}A, telecom light is coupled from a fiber into the TM$_{00}$ mode in the waveguide through an infrared GC (labeled as GC1 in Fig.~\ref{fig2}A, microscopic photo shown as Fig.~\ref{fig2}B), which reaches a peak coupling efficiency of about 24\% near 1550~nm with a 3-dB bandwidth larger than 70~nm, and about 23\% at 1560~nm (Fig.~\ref{fig2}F). The waveguide is folded into a paper-clip geometry whose three straight sections are periodically poled~\cite{yu_Efficient_2025} for quasi-phase-matched SHG, as resolved by scanning electron microscopy of the poled domains in Fig.~\ref{fig2}C. With a poling period of 2.36~$\mu$m, quasi-phase-matching converts the infrared TM$_{00}$ mode near 1560~nm (left inset of Fig.~\ref{fig2}A) to the visible TM$_{00}$ mode near 780~nm (right inset of Fig.~\ref{fig2}A).

(ii) \textit{Emission}. The 780~nm light is then emitted vertically into free space by a visible-light GC with a uniform filling factor (labeled as GC2, Fig.~\ref{fig2}D). Modeling the emission as a Gaussian beam with its waist at the chip surface and a waist size set by the grating-coupler aperture (about 20~$\mu$m in $x$ and 30~$\mu$m in $y$) predicts a beam of 1.5~mm ($x$) by 0.9~mm ($y$) after 30~mm of propagation. The measured beam (Fig.~\ref{fig2}H) is an ellipse of 2.8~mm (long axis) by 1.6~mm (short axis), larger than predicted. We attribute this to light being scattered by only part of the GC, which reduces the effective waist and increases the divergence. Figure~\ref{fig2}I shows the emitted SH power of one SHG waveguide versus pump wavelength for 230~mW of on-chip infrared power (1~Watt off-chip power). Film-thickness inhomogeneity and imperfect constructive interference among the three poled sections make the spectrum deviate from an ideal sinc$^2$ profile and appear irregular. The spectrum blue-shifts at about 0.5~nm/$^{\circ}$C with increasing chip temperature. At about 42.5~$^{\circ}$C, one peak is tuned to a pump wavelength of 1560.48~nm (780.24~nm SH output), where the emitted 780.24~nm power reaches about 0.9~mW.

(iii) \textit{Recollection}. The passive waveguide recollects the returned probe and routes it off chip. It has the same cross-section as the SHG waveguide but is not poled, serving only as a 780~nm routing channel, and its two ends use visible-light GCs of different design. The input coupler (labeled as GC3 at right side of Fig.~\ref{fig2}A) recollects the probe reflected from the Rb system and uses the same design as the SHG-waveguide emission coupler, ensuring a matched mode for the returning beam. The output coupler (labeled as GC4 at left side of Fig.~\ref{fig2}A, microscopic image in Fig.~\ref{fig2}E) is instead optimized for fiber coupling, with a smaller aperture (about 25~$\mu$m in $y$ by 20~$\mu$m in $x$) and a filling factor that varies gradually along the propagation direction to shape the beam close to the fiber-core size. Figure~\ref{fig2}G shows the transmission of a pair of these small fiber-coupling couplers joined by a 250~$\mu$m waveguide, measured at 16.6\%. We note that the SMF-28 fiber used here supports multiple modes at 780~nm, and the connecting waveguide supports at least six modes, so the spectrum in Fig.~\ref{fig2}G exhibits multimode interference and does not permit an accurate extraction of the single-coupler efficiency. Square root of the measured transmission gives an rough 40\% coupling efficiencyfor the GC4. Fig.~\ref{fig2}H shows the measured end-to-end transmission through the passive waveguide is about 9.1\% at 780~nm. Deduct out the coupling efficiency for the GC4 and the propagation loss in the 5~mm-long waveguide, we estimate the visible-band GC3 coupling efficiency as roughly 38\%.

Having characterized the chip, we now integrate it with the Rb atomic system to close the re-entrant loop and interrogate the atomic reference. Figure~\ref{fig3}A shows the experimental setup. A tunable laser input at 1560.48~nm is injected into an SHG waveguide after passing through an erbium-doped fiber amplifier (EDFA) and a polarization controller (PC). On the left side of the chip, the input couplers of the SHG waveguides and the output couplers of the passive waveguides form a 250~$\mu$m-pitch-period array that is matched to a standard fiber array with total-internal-reflection end faces (Fig.~\ref{fig3}B). The generated 780.24~nm light is emitted vertically from the chip and passes through the Rb vapor cell as the pump beam. The cylindrical Rb cell (height $\sim$30~mm) is mounted in a Teflon holder with a 7.5~mm-diameter clear aperture and stabilized at 48~$^{\circ}$C to provide a suitable vapor density for SAS. Above the cell, a convex lens ($f = 50$~mm) collimates the emitted beam, and a beam splitter reflects about 98.6\% of the light back along the same optical path, so that the retro-reflected beam passes through the cell a second time as the probe. By tuning the beam-splitter angle, the reflected probe is redirected onto the passive-waveguide input coupler and recollected on chip. The recollected probe then propagates through the passive waveguide, exits via the fiber array, and is detected by a photodetector (PD). A concave lens ($f = -50$~mm) and a microscope camera above the beam splitter image the emitting coupler and the reflected spot to assist alignment (not plotted in Fig.~\ref{fig3}A).

An additional requirement of the re-entrant loop for the SAS is that the forward and backward beams overlap within the cell. The pitch between the emitting and recollection grating couplers is 957~$\mu$m, so the angle between the pump and probe paths is below 1$^{\circ}$, yielding a beam-center separation of only hundreds of micrometers at the cell, well below the millimeter-scale beam diameter and thus sufficient for spatial overlap. Equally important is the optical intensity delivered to the atoms: the 0.9~mW of emitted 780.24~nm light, spread over the measured 2.8~mm by 1.6~mm beam, corresponds to a peak intensity of several mW/cm$^2$, comparable to or above the saturation intensity of the $^{87}$Rb $D_2$ line ($I_s \approx 1.6$~mW/cm$^2$), and gives a power-broadening of approximately 25~MHz, which is relatively smaller than that in Ref.~\cite{hummon_Photonic_2018}. The chip therefore delivers enough on-chip-generated power to saturate the transition, which is what allows the sub-Doppler Lamb dips, rather than only the Doppler-broadened absorption, to be resolved.

As shown in Fig.~\ref{fig3}C, the PD records a maximum recollected probe power of about 0.7~$\mu$W at off-resonance frequency. This implies re-entrant probe power of roughly 2.9~$\mu$W on chip and the recollection efficiency of approximately 0.32\%, which is mainly due to beam distortion in the Rb interaction region, reflection and scattering at the cell walls, and residual alignment mismatch between the grating couplers and either the fiber array or the reflected probe beam. Despite this low recollection efficiency, the recollected power is more than sufficient for spectroscopy: as shown in Fig.~\ref{fig3}D, the probe power fluctuates by a 0.1~$\mu$W standard deviation over 100 sampling rounds, so that the sub-Doppler features and the derived error signal are limited by the atomic response rather than by the stability of the re-entrant link. These results confirm that the recollected signal is stable enough for spectroscopy and further frequency locking to the atomic transitions.

Scanning the laser frequency, we measure the SAS of $^{87}$Rb (Fig.~\ref{fig3}D). By adjusting the vapor cell temperature to control the Rb atom density, we optimize the signal-to-noise ratio. At 48.0$^{\circ}$C, we can clearly resolve the hyperfine features of the $^{87}$Rb D2 line. Using one of the narrow features, the crossover resonance between the $F=2\rightarrow F'=2$ and $F=2\rightarrow F'=3$ transitions (inset of Fig.~\ref{fig3}D), as a frequency reference, we perform Pound-Drever-Hall (PDH) locking for 2~hours. Fig.~\ref{fig3}(E) shows that the frequency drift of the infrared laser, derived from the PDH error signal, is suppressed to within $\pm$280~kHz, compared with a free-running drift of 190~MHz over 2~hours. Crucially, this stabilization is achieved by addressing the atomic transition directly within the PIC: apart from the cell, a collimating lens, and a retro-reflecting beam splitter, the generation, routing, and detection are all performed on chip, with no free-space frequency-conversion optics. In this sense, the saturated-absorption reference is folded into the on-chip optical path itself, so that in principle a Rb cell sealed together with a reflector could be bonded directly onto the chip as a self-contained, re-entrant atomic-frequency module.

\smallskip{}
\noindent \textbf{\large{}{}Discussion}{\large\par}
\noindent In summary, we have demonstrated a re-entrant chip-free-space photonic interface that embeds a narrow atomic transition directly into a PIC: telecom light is frequency-doubled, emitted through a $^{87}$Rb vapor cell, recollected on the same chip, and used to lock the laser to the atomic resonance within $\pm 280$~kHz over 2~hours. In effect, the atomic internal states become an addressable functional block of a photonic system-on-chip~\cite{Grosman2025}. Because the interface accesses the atomic internal states, the same architecture extends naturally to atomic internal-state-based sensing~\cite{Brown2025,Kitching2025,Ma2026}, quantum memories~\cite{Glorieux2023,Finkelstein2023}, modular quantum-network nodes~\cite{Kimble2008}, and portable optical atomic clocks~\cite{Newman2019}. Moreover, replacing the $^{87}$Rb cell with other atomic or molecular vapors turns the platform into a chip-scale reference for integrated frequency-comb spectroscopy~\cite{Gaeta2019}.

Looking forward, the re-entrant topology allows the returned signal to re-enter the full on-chip toolbox, such as integrated filtering, interferometry, coherent detection, multiplexing, and feedback, rather than terminating in free space. One critical engineering objective is to improve the reflection-and-recollection efficiency, which is dominated not by the architecture itself but by beam distortion in the interaction region, reflection and scattering at the cell walls, and residual alignment mismatch. This budget can be substantially improved by replacing the passive mirror with an engineered phase-compensating reflector or programmable metasurface~\cite{Moitra2023} that maximizes the mode overlap of the returning beam with the recollection GC, by anti-reflection treatment of the cell interfaces, and by co-designing the emission and recollection couplers for a matched retro-reflected mode. Beyond gaseous media, the same emit-interact-recollect cycle can address solid gain media~\cite{Tran2022}, condensed samples, and biological analytes~\cite{Estevez2012} on a single chip: an embedded gain medium would enable chip-scale optical amplifiers and lasers in bands that are difficult to integrate monolithically, while thin analyte layers would enable on-chip absorption, phase, or fluorescence readout.

Taken together, these directions point to a broader integration paradigm. Because each external medium is held in a self-contained three-dimensional module that couples to the chip only through a standardized GC-mirror-GC footprint, such modules can in principle be placed onto a photonic circuit in much the same way that capacitors and other discrete components are soldered onto an electronic circuit board. The millimeter-scale span and relaxed alignment tolerances demonstrated here make this pick-and-place picture concrete, allowing a library of interchangeable functional cubes, for atomic references, gas or biological sensing, optical gain, or quantum memory, to be mounted, replaced, and reworked on a common active photonic motherboard. In this view, the re-entrant interface provides a general principle for the heterogeneous integration of otherwise non-integrable matter into PICs, contributing a concrete route toward truly functional photonic systems-on-chip.

\section*{Acknowledgments}
{This work was supported by the Strategic Priority Research Program of the Chinese Academy of Sciences (XDA0520503), National Natural Science Foundation of China (12273087,62305214, 92576113,12504454, and 12293053).} {This work was also supported by Quantum Science and Technology-National Science and Technology Major Project (2021ZD0300701), Industrial Prospect and Key Core Technology Projects of Jiangsu Provincial Key R\&D Program (BE2022071), Shanghai Science and Technology Committee (25JD1405800), the Fundamental Research Funds for the Central Universities, and USTC Research Funds of the Double First-Class Initiative.} The numerical calculations in this paper were performed on the supercomputing system in the Supercomputing Center of USTC, and this work was partially carried out at the USTC Center for Micro and Nanoscale Research and Fabrication and the ShanghaiTech Material Device Lab (SMDL).

\end{document}